\documentclass[prl,reprint,superscriptaddress]{revtex4-1}
\usepackage{graphicx}

\begin{document}

\title{High-Speed Tunable Microcavities Coupled to Rare-Earth Quantum Emitters}%

\author{Kangwei Xia}%
\email[Kangwei Xia:]{kangwei.xia@pi3.uni-stuttgart.de}
\affiliation{3. Physikalisches Institut, University of Stuttgart, 70569 Stuttgart, Germany}

\author{Fiammetta Sardi}%
\email[Fiammetta Sardi:]{f.sardi@pi3.uni-stuttgart.de\\These authors contributed equally: Kangwei Xia, Fiammetta Sardi}
\affiliation{3. Physikalisches Institut, University of Stuttgart, 70569 Stuttgart, Germany}

\author{Colin Sauerzapf}%
\affiliation{3. Physikalisches Institut, University of Stuttgart, 70569 Stuttgart, Germany}

\author{Thomas Kornher}%
\affiliation{3. Physikalisches Institut, University of Stuttgart, 70569 Stuttgart, Germany}

\author{Hans-Werner Becker}%
\affiliation{RUBION, Ruhr-Universität Bochum, 44780 Bochum, Germany}

\author{Zsolt Kis}%
\affiliation{Wigner Research Center for Physics, Institute for Solid State Physics and Optics, H-1121 Budapest, Hungary}

\author{Laszlo Kovacs}%
\affiliation{Wigner Research Center for Physics, Institute for Solid State Physics and Optics, H-1121 Budapest, Hungary}

\author{Roman Kolesov}%
\affiliation{3. Physikalisches Institut, University of Stuttgart, 70569 Stuttgart, Germany}

\author{J\"org Wrachtrup}%
\affiliation{3. Physikalisches Institut, University of Stuttgart, 70569 Stuttgart, Germany}
\affiliation{Max Planck Institute for Solid State Research, 70569 Stuttgart, Germany}

\date{\today}%

\begin{abstract}

Electro-optical control of on-chip photonic devices is an essential tool for efficient integrated photonics. Lithium niobate on insulator (LNOI) is an emerging platform for on-chip photonics due to its large electro-optic coefficient and high nonlinearity$^{[1]}$. Integrating quantum emitters into LNOI would extend their versatile use in classic photonics to quantum computing and communication$^{[2,3]}$. Here, we incorporate single rare-earth ions (REI) quantum emitters in electro-optical tunable lithium niobite (LN) thin films and demonstrate control of LN microcavities coupled to REI over a frequency range of 160 GHz with 5 µs switching speed. Dynamical control of the cavities enables the modulation of the Purcell enhancement of the REIs with short time constants. Using the Purcell enhancement, we show evidence of detecting single Yb$\mathrm{^{3+}}$ ions in LN cavities. Coupling quantum emitters in fast tunable photonic devices is an efficient method to shape the waveform of the emitter$^{[4]}$. It also offers a platform to encode quantum information in the integration of a spectral-temporal-spatial domain to achieve high levels of channel multiplexing, as well as an approach to generate deterministic single-photon sources$^{[5,6]}$.
\end{abstract}
\maketitle

Integrated optics is of paramount importance, especially for modern optical information technology. It enables versatile routing of photons, wavelength conversion and in combination with electro-optical material, an interface to electronic components. In this regard, lithium niobate (LN) is a key optical material due to its broad transmission window, large electro-optic coefficients and nonlinearity$^{[7]}$, with various applications, such as electro-optical modulators, coherent photon conversion and quantum memories$^{[8,9]}$. The recent commercialization of lithium niobate on insulator (LNOI) offers an attractive platform to highly integrate photonic devices on-chip$^{[1]}$. The low operation voltage and fast electro-optical modulation speed result in a variety of attractive applications, ranging from broadband frequency comb generation$^{[11]}$ and fast electro-optic modulators$^{[12,13]}$ to high-quality factor cavities$^{[14]}$, micro-lasers$^{[15]}$ and more. However, LNOI has not shown similar functionality in integrated quantum technology. While it has been used as classic electro-optical element in optical quantum technology, such as photon pair sources and single photon wavelength converters$^{[16]}$, a direct quantum functionalization in e.g. quantum nano mechanics or via doping with quantum emitters has not yet been achieved. Specifically, the latter method would allow for the realization of photonically interfaced quantum nodes with the ability to dynamically control quantum emitters$^{[17-19]}$. Such a versatile control of individual qubits in the spectral, temporal and spatial domain respectively, would further facilitate  construction of the scalable multi-qubit quantum networks$^{[5,6]}$.
  
  \begin{figure}
\includegraphics[width=3.5in]{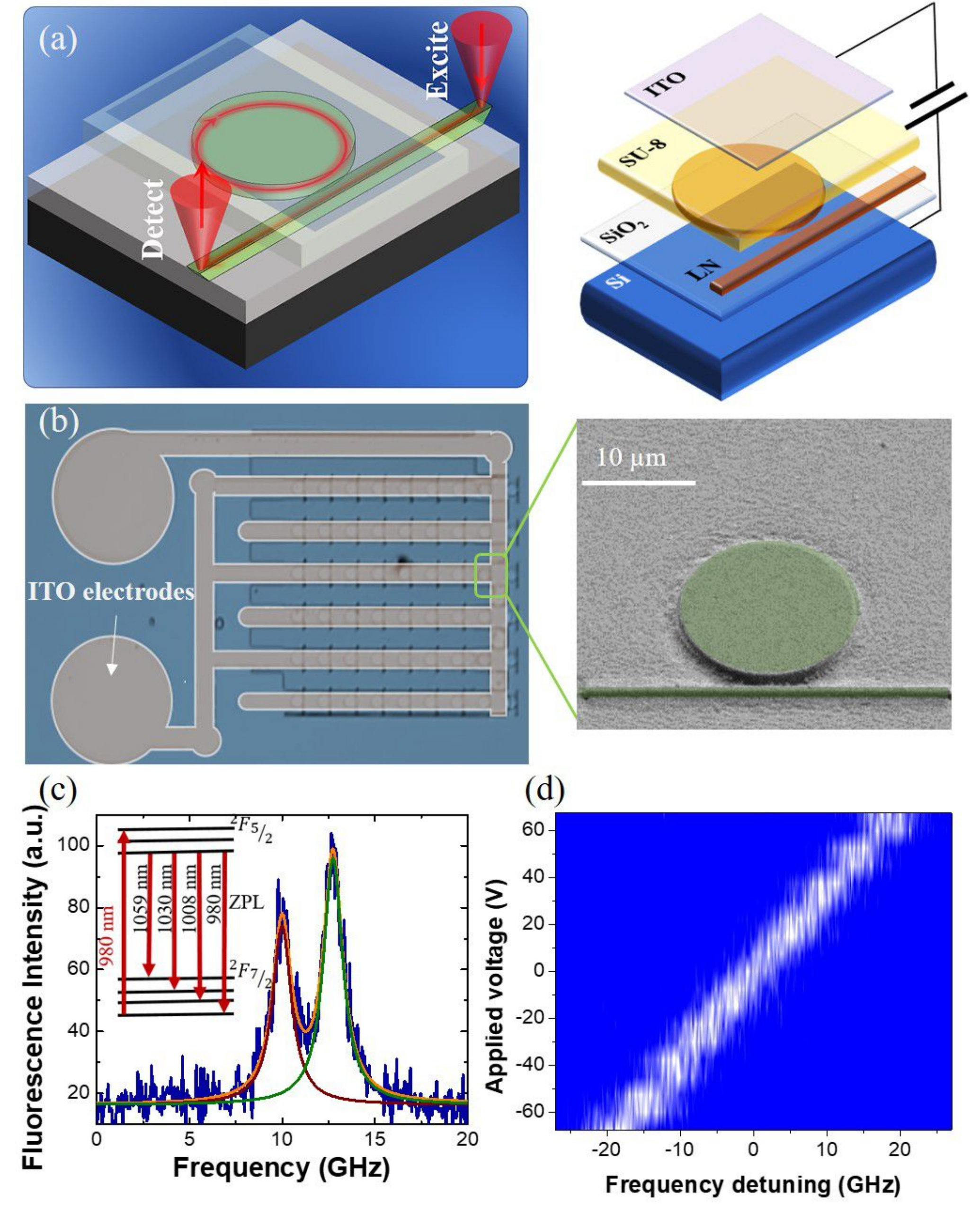}%
\caption{ \textbf{Optical and electro-optical characterization of LN microdisks.} a) Schematics of our device. On a layer of amorphous SiO$_2$ a microdisk with a waveguide is fabricated by e-beam lithography and etching. Photonic structures are capped with a layer of SU-8 on which we deposited a 30 nm layer of ITO. The right graph is an explored view of the device design. b) Optical images of the fabricated microdisks on a LN thin film with top ITO electrodes. The right graph is a SEM image of the microdisk. Incarnadine represents the ITO electrodes and the green pattern shows the LN microdisk with coupling waveguide. c) Linewidth of one of the monolithic LN microdisk fabricated on-chip with a Q-factor of 244,400. Insert shows the energy level of Yb:LN. d) Electro-optic tuning of cavity mode depending on the bias voltage applied to the device with a coefficient of 270 MHz/V. \label{figure1}}
\end{figure}  

One option to achieve quantum-functionalized LNOI is to dope the material with rare earth ions (REI)$^{[20-22]}$. REIs in solids have recently emerged as promising systems. They offer advantages in both optical and spin aspects: various emission wavelengths including telecom band compatible with fiber quantum networks, spin transitions with long coherence time that enables long lasting quantum memory$^{[23-25]}$. However, the 4$f$-4$f$ dipole forbidden optical transitions of REIs make optical readout at single ion level challenging$^{[26]}$. The incorporation of REIs into photonic cavities is an efficient approach to enhance quantum light-matter interface, where the lifetime of REIs reduces due to the Purcell effect$^{[27],[17]}$. Currently, the coupling of single REIs to photonic devices is based on evanescent tail coupling between  photonic devices and bulk YSO crystals or on direct fabrication of nanophotonic devices in YVO$\mathrm{_4}$ bulk crystals$^{[28,19]}$. This is mostly due to the fact that it has proven to be challenging to observe optically active single REIs in electro-optical thin LNOI films. Typically, the incorporation of REIs into LNOI requires a smart cut of the REI-doped LN bulk crystal$^{[20,21]}$. Another approach is directly doping REIs in LN thin films by ion implantation$^{[30]}$. The implantation is followed by high-temperature annealing, required to stabilize the dopants in the correct lattice sites and heal the lattice damage. However, this process forms cracks in LNOI and substantially reduces the optical quality of the material.

Here, we directly doped REIs in LN thin films by means of ion implantation. We perform hole-burning spectroscopy of Yb-implanted LNOI and show that post-annealing at 650~$^{\circ}$C is sufficient to achieve spectrally narrow excitation of Yb$\mathrm{^{3+}}$ ions, as well as excellent charge stability. This procedure is also compatible with fabricating photonic elements, like electro-optically tunable high-Q microcavities. With these elements, we are able to modulate the Purcell effect of Yb$\mathrm{^{3+}}$ ions  dynamically.  By coupling  the REIs to the cavity and hence increasing their emission rate, we provide evidence of detecting single Yb$^{3+}$ ions.

In  experiments, we implant Yb ions with an energy of 1.5 MeV and a fluence of 10$\mathrm{^{12}}$ ions/cm$\mathrm{^{2}}$ into LN films. After ion implantation, we further post-anneal the Yb-implanted LN thin film at 650 $^{\circ}$C (see SI Note 2) without observing any damage to the film. The annealing temperature is still 500~°C lower than required for Yb doping of LN during crystal growth. Thus, before using the REI-implanted thin film, further characterization is needed to evaluate their optical properties. We perform hole burning spectroscopy (HBS) to assess the quality of the implanted Yb ions in the LN waveguide and Yb-doped bulk LN crystal at cryogenic temperatures (the experimental details are described in SI Note 4). Yb-doped LN bulk crystals show spectral hole widths of 2$\mathrm{\pi}$×58~MHz, while the Yb-implanted LN long waveguide shows 2$\mathrm{\pi}$×82~MHz (details see SI Note 4). The results suggest that post-annealing at 650 $^{\circ}$C is sufficient to activate and stabilize the implanted Yb ions in LN thin films, showing a similar optical performance to the one in a bulk crystal. This gives us a simple approach to directly incorporate REIs in pre-fabricated photonic devices, based on commercially available LN thin film. 

  \begin{figure*}
\includegraphics[width=7in]{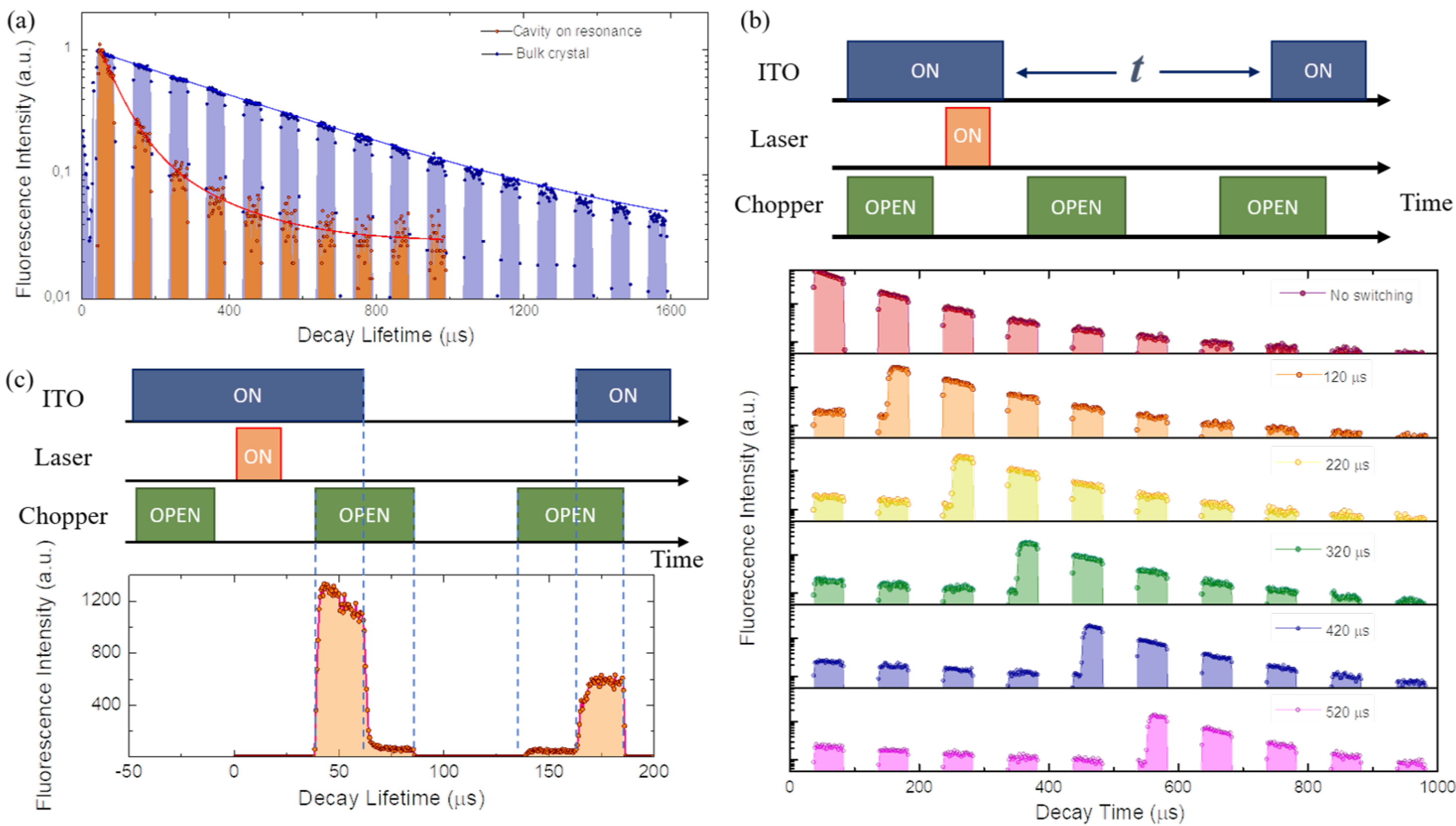}%
\caption{ \textbf{Yb$\mathrm{^{3+}}$ in the LN microcavity.} a) Lifetime of implanted Yb ions on- and out of resonance with the cavity. The blue curve shows a long decay time of 430 µs for off-resonance ions and the red curve shows a shorted on-resonance decay lifetime. b) Schematics of the fast tuning protocol, showing how the cavity stores energy, when not on-resonance with Yb ions. c) High resolution measurements to show the time constant of our device for frequency tunig. Switching is shown to occur on the order of 5 µs, mainly limitated by electronics.  \label{figure2}}
\end{figure*}

We further shape the Yb-implanted LN thin film to microcavities (the fabrication protocol is described in Methods and SI Note 3). A scanning electron microscope (SEM) image of one of the cavities is shown in Fig. 1 (b). The microdisk resonator has a size of 7 µm radius. A waveguide positioned next to the microdisk evanescently couples light into and out of the cavity. To enhance its output efficiency, a 45° undercut is milled at the ends of the waveguide (see SI Note 3). The typical Q-factor of the  fabricated cavities is in a range of 50,000-150,000 with a mode volume $\mathrm{V_{mode}=50(\lambda/n)^3}$. The highest Q-factor of the cavity is 244,730, as shown in Fig.~1(c). According to simulations, the Q-factor of the microdisk can exceed 10$^9$ when only radiative loss is taken into account (see SI Note 3). The Q-factor of the current cavities is mainly limited by scattering losses that can be further improved by optimizing the fabrication protocol.

To achieve electro-optic tunability, an additional layer of indium tin oxide (ITO) is deposited on the LN cavity serving as the top electrode, with a negative 1 µm photoresist (SU-8) serving as a spacer. The Si substrate builds the bottom electrode. By applying a bias voltage between the electrodes, the reflective index of the LN cavity is modulated, resulting in the tuning of the cavity resonance frequency, as shown in Fig. 1(d). We further estimate the device's electro-optic coupling by applying ±60 V external voltage with a step width of 15~V. The response of the cavity turns out to be linear, indicating the absence of hysteresis from our device, with a coefficient of 270~MHz/V. In the experiment, a maximum voltage of $\pm$300~V is applied corresponding to the tuning range of $\pm$81~GHz.

With  electro-optical tuning, the cavity frequency can be brought in  and out of resonance with Yb$\mathrm{^{3+}}$:LN. For cavity QED experiments, a microdisk with a Q-factor of 79,833 and high coupling between the cavity and the waveguide is further investigated. When the frequency of the cavity is tuned on resonance with the ZPL of Yb$\mathrm{^{3+}}$:LN, we observe a shortening of the Yb$\mathrm{^{3+}}$ lifetime in the cavity, as shown in Fig. 2 (a). A double-exponential-decay-function yields a lifetime of 51$\pm$4~µs and 182$\pm$40~µs. The different shortening of the lifetime suggests that Yb$\mathrm{^{3+}}$ ions couple to the cavity with different strengths because of their different positions in the cavity.  The lifetime in bulk Yb:LN is 430$\pm$2~µs. Hence Yb$\mathrm{^{3+}}$ coupling to the cavity results in a Purcell factor of 7.6. The lifetime shortening is three times lower than the theoretical estimation, based on Q/V ratio. This is presumably caused by inappropriate Yb dipole orientation in the $z$ cut LNOI.

  \begin{figure*}
\includegraphics[width=7in]{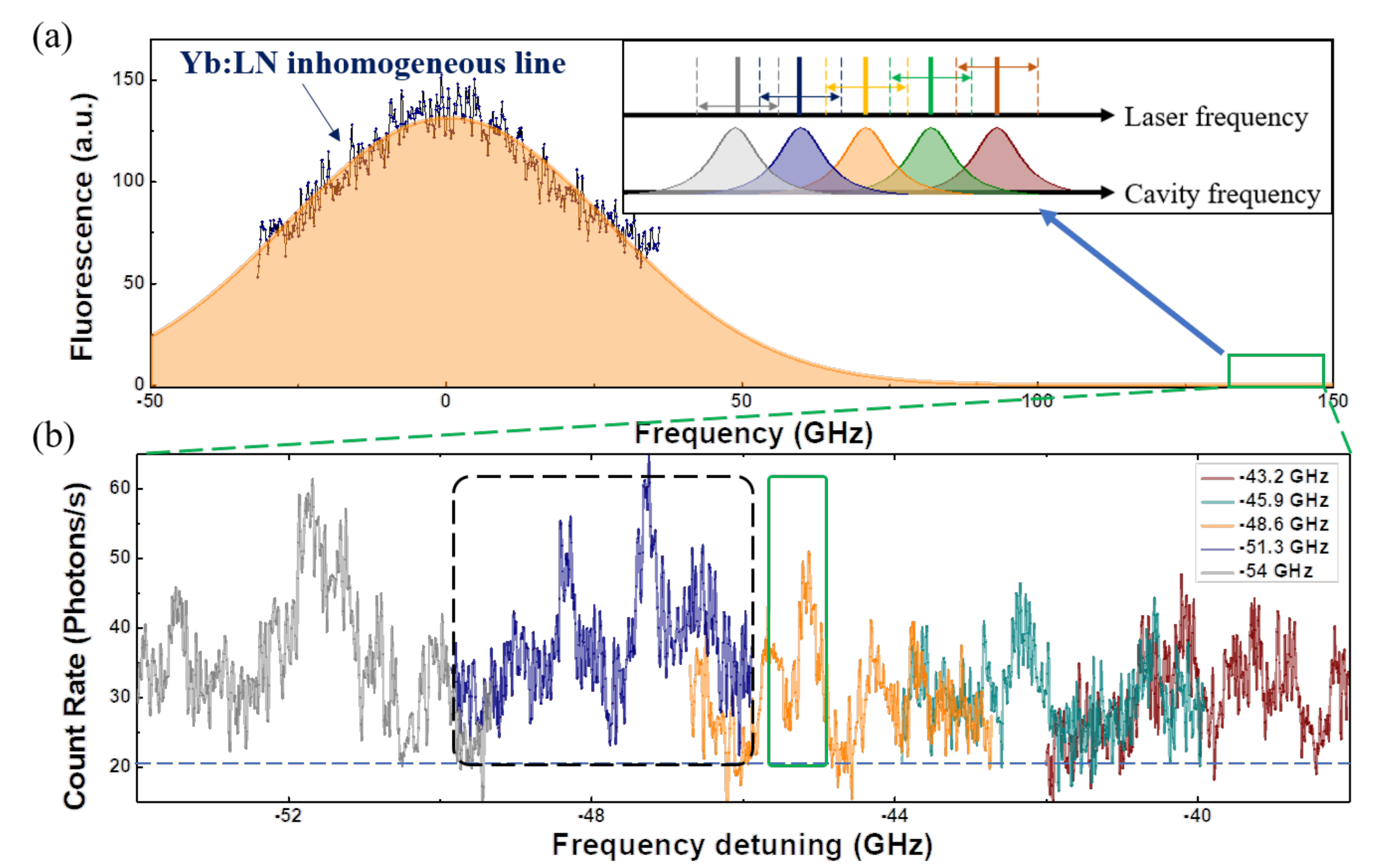}%
\caption{ \textbf{Evidence of detecting single Yb ions in LN.} a) Inhomogeneous line of implanted Yb$\mathrm{^{3+}}$ in LN with a linewidth of 2$\mathrm{\pi}$×64±0.6 GHz.  The green rectangle marks the position where the PLE spectra of b) are taken. The inserted graph is a schematic of the experimental protocol. The frequency of the cavity is fixed to a certain position. The laser frequency is swept around the resonance frequency of the cavity. b) PLE spectrum recorded at the tail of the inhomogeneous line profile of Yb implanted in LN marked. A cavity with a Q-factor of 64,065 is optically tuned from the center of the spectrum to the tail with a step-width of 2.7 GHz and a total tuning range of 10 GHz. The discrete peaks have an average linewidth of 2$\mathrm{\pi}$×247±10 MHz. Overlapping of peaks shows that the cavity can be electro-optically tuned back and forth across the spectrum. \label{figure3}}
\end{figure*}  

In the following, fast electro-optic dynamic control of the microcavity coupled to REIs is demonstrated: (1) The frequency of the cavity is tuned in resonance with the Yb ions. (2) A laser pulse is generated to resonantly pump the Yb ions to the excited states. (3) The excited Yb ions start to decay back to their ground state with 50 µs lifetime due to the cavity QED. (4) Before the Yb ions completely decay to the ground state, the cavity resonance frequency is tuned off-resonance (54 GHz frequency detuning) from the excited Yb ions, thus prolonging their lifetime from 50 µs  to 430 µs. Since this emission from Yb ions coupled to the cavity is now off-resonance, the excitation is stored in the cavity and is not coupled to the waveguide (5). After a certain waiting time, the cavity is tuned back on resonance with the excited Yb ions, resulting in the restoration of fast decay of the Yb ions. The experiment has been carried out with different waiting times, as shown in Fig. 2(b). 

To determine the electro-optical switching time of the microcavity, measurements with fast time resolution are performed (see Fig.~2c). According to the rising and falling slopes of the fluorescence signals, the switching time of the microcavity is $\sim$ 5 µs, limited by the bandwidth of the high-voltage operational amplifier in use.

We further investigated the detection of a single ion, based on the increased fluorescence signal due to Purcell enhancement. In the experiment, a cavity with a Q factor of 64,065 is used. Instead of at the center of the inhomogeneous line of Yb$\mathrm{^{3+}}$, the cavity resonance frequency is tuned along the tail of the inhomogeneous line of 980~nm Yb$\mathrm{^{3+}}$ optical transition (more than 140~GHz away from the center of the inhomogeneous line, as shown in  Fig.~3 (a)). A laser power of 22 nW is applied to excite single Yb ions. The respective photoluminescence excitation spectrum (PLE) is shown in Fig.~3(b), where discrete lines are observed. In total, 24 discrete lines are fitted with an average linewidth of $\mathrm{2\pi\times247\pm9~MHz}$ (SI Note 5). We further plot one of the peaks, as shown in Fig.~4, with a fitted linewidth of 2$\mathrm{\pi}$×215±22~MHz. The linewidth is twice as large as the spectral hole width that we obtained by hole-burning spectroscopy of Yb$\mathrm{^{3+}}$ implanted in a LN waveguide (see Fig.~4 (a)). We suspect that the broadening is caused by spectral diffusion. 

  \begin{figure}
\includegraphics[width=3.5in]{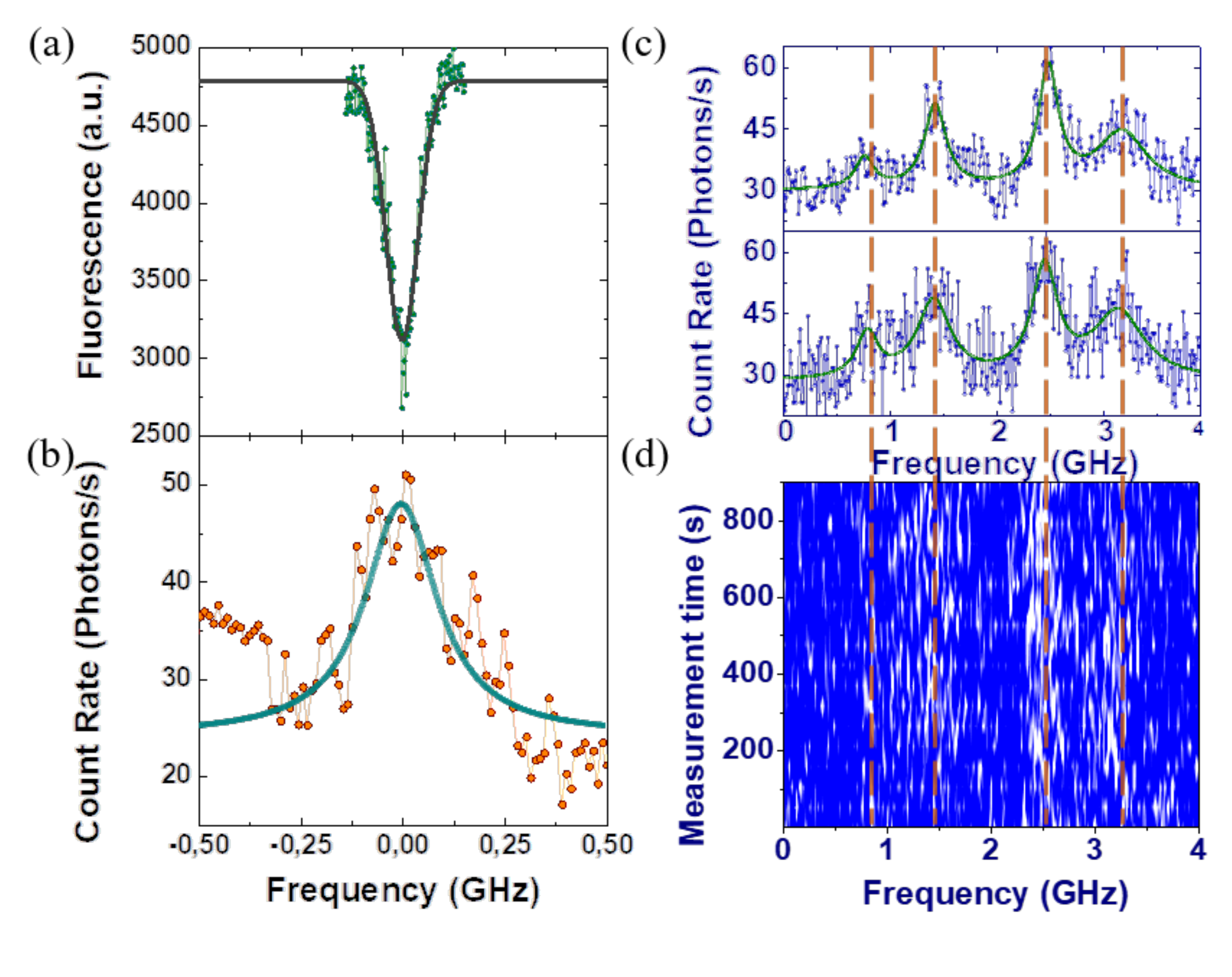}%
\caption{ \textbf{Spectroscopy of single Yb ions in LN.} a) Hole burning spectroscopy of Yb implanted LN  waveguide with a spectral hole width of 2$\mathrm{\pi}$×97± 2.3 MHz. b) Plot of an individual PLE peak marked as a rectangle with green solid lines in Fig.3 (b). The linewidth is 2$\mathrm{\pi}$×215± 22 MHz. c) Single PLE spectrum of a sweep marked as a rectangle with black dash lines in Fig. 3 (a). The sweep is from lower frequency to higher frequency (upper curve) and reverse (lower curve). Both spectra show  identical features, demonstrating that discrete peaks are not experimental artefacts but reproducible spectral features. d) Acquisition of PLE spectra for more than 800 s over the same range as c) showing no obvious spectral diffusion of ions inside the cavity \label{figure4}}
\end{figure}  

Another important feature of emitters coupled to cavities is their spectral stability. Often, the fabrication of optical cavities results in spectrally unstable  emitters. We investigate the spectral stability by observing consecutive PLE measurements over an extended period of time (Fig.~4(d)). Remarkably, we do not observe a strong spectral diffusion. Moreover, due to the flexible electro-optical tunability, we can detune the cavity resonance frequency in steps of 2.7 GHz and obtain the PLE spectra in different ranges, as shown in Fig.~3(b). The total sweeping range is ~10 GHz, which is two times broader than the intrinsic linewidth of the cavity.  

According to the PLE measurements, the count rate of the discrete peaks is 20-30 counts/s, which is consistent with our estimation (see Methods). This suggests that the discrete peaks are corresponding to single Yb ions in the LNOI microcavity. Due to the low count rate, we are not able to perform e.g. second-order correlation measurements to unambiguously support our assumption.

\subsection{Conclusion}

In conclusion, we incorporate REI quantum emitters into LN thin film with optical properties similar to native REIs. The electro-optic control of the LN microcavities enables couplings of REIs to be controlled at 5 µs switching speed and over a 160 GHz  frequency range. Facilitated by the Purcell enhancement, we further observe single Yb$\mathrm{^{3+}}$ ions in the cavities. The demonstration of dynamic control of the on-chip photonic devices enables the shaping of a single photon waveform. The spectral, temporal and spatial selectives of the single ions coupled to the cavities will be a flexible and scalable approach to encode quantum information by employing high levels of channel multiplexing. Controlling the cavities in  and out of the resonance of single ions will be a highly desirable method to generate deterministic single photon sources. Furthermore, LN thin films and an additional spin-free material, such as a YSO thin film, can form a hybrid quantum system. The long-lived quantum memory can be dynamical controlled by the electro-optical tuning of the LN photonic devices, which enables  spectral multiplexing of quantum memories.

\subsection{Sample preparation}
After the implantation and annealing, the samples are further cleaned in a Piranha solution to then evaporate a 30 nm Titanium layer through E-gun evaporation. Subsequently a double layer of PMMA(400 nm and 80 nm) is spin-coated directly onto the sample. The e-beam lithography is then performed with a RAITH E-line machine. A layer of 120 nm Nickel/Chromium alloy with a percentage of 80 to 20 is evaporated using e-gun and specific structure are created by lift-off procedure in NEP at 90 $^{\circ}$C. Micro-cavities are then transferred to the LNOI by Plasma etching in a Plasma Oxford Instruments machine with a recipe consisting of different steps. In the first part SF$_6$ gas is utilized in order to have anisotropic plasma and optimize the transfer in LN with help of smoothing leverage of no ICP AR milling. Then a hard mask of NiCr and Ti are wet etched with the use of Nitric acid and Piranha solution. Lastly, the last 70 nm of LN are removed by Ar milling, leaving a 10 nm of thin film before SiO$_2$ and Si chip.

\subsection{Single Yb$^{3+}$ ions’ photon count rate estimation}

According to the Purcell enhanced lifetime we measured, the count rate of single Yb$^{3+}$ in LN cavity is calculated to be 9,000 photon/s. The total amount of the detected photons is estimated to be:

$9000\times\eta_c (30\%)\times\eta_o (30\%)\times\eta_e (50\%)\times\eta_m (70\%)\times\eta_{chopper} (50\%)\times\eta_d (60\%)$
=40 counts/s,

where $\eta_c$ is the coupling efficiency between the microcavity to the waveguide; $\eta_o$ represents the output efficiency at the end of the waveguide; $\eta_e$ indicates the signal collected from only one end of the waveguide; $\eta_{m}$ is the collection and transmission efficiency of the optical microscopy;  $\eta_{chopper}$ is the duty cycle of the chopper; and $\eta_{d}$ is the efficiency of the superconducting single photon detector.

\subsection{Quality of  micro-cavities}
      For the cavity, we studied in Fig.~2 (and Fig.~3), the linewidth is 3.83~GHz (and 4.9~GHz); the Q factor is 79,833 (and 64,065); the REI lifetime reduction is measured to be 50.0 us (and 62.7 µs). The Purcell factor is C= $\gamma_0$/ $\gamma_c$-1=7.6 (and 5.9), where $\gamma_c$, $\gamma_0$ represent the decay rates of Yb:LNOI coupled to the cavity and the natural lifetime of the bulk Yb:LN respectively. These measurements further allow us to estimate the QED parameters for our cavities as ($g$,$\kappa$, $\Gamma_c$ )=2
     $\mathrm{\pi}$×(1.64 MHz,3.83 GHz,370 Hz) and (2 $\mathrm{\pi}$ ×(1.62 MHz,4.9 GHz,370 Hz)), where $g$,$\kappa$ are related to the single photon Rabi frequency and cavity decay rate. 
      We further estimated for two cavities, 88\% (85.6\%) of the Yb:LNOI fluorescence is  emitted into the cavity mode and the branching ratio of emission into 980 nm transition is improved from 25\% to 91.5\% (89.1\%).

\subsection{Acknowledgments}
We would like to thank Arkadiy Kolesov and Frank Thiele for their support with the experiment, and Dr. Matthias Widmann for helping with artwork design. We would like to thank the fruitful discussion with Dr. Qi-chao Sun, Prof. Kai-Mei Fu, and Dr. Philippe Goldner. We would like to thank Sabrina Jenne for language corrections. R. K. acknowledges financial support by the DFG (Grant No. KO4999/3-1), and R. K. and J. W. acknowledge financial support by the FET-Flagship Project SQUARE, the EU via SMeL and QIA, as well as the DFG via FOR 2724 and GRK 2642. J. W. acknowledges BMBF, Q. Link. X.  Z. K. and L. K. acknowledges the financial support by the National Research, Development and Innovation Fund of Hungary within the Quantum Technology National Excellence Program (Project Nr. 2017-1.2.1-NKP-2017-00001) .

\subsection{Author contributions}
K. X. and F. S. contribute equally. K. X., F. S, R. K. and J. W. conceived and designed the experiments. F. S., R. K., K. X., T. K. fabricated the microstructures. H. B., Z. K. and L. K. provided the samples. K. X., F. S., C. S. and R. K. built the experimental setups.K. X., F. S., C. S. and R. K. performed the experiment and analysed the data. K. X., F. S., C. S., T. K., R. K., and J. W. wrote the manuscript with input from all authors and R. K. and J. W. supervised the whole project.


\subsection{Reference}
[1]          A. Boes, B. Corcoran, L. Chang, J. Bowers, A. Mitchell, Status and potential of lithium niobate on insulator (LNOI) for photonic integrated circuits, Laser Photonics Rev. 12 1700256, 2018. 

[2] 	D. Awschalom, R. Hanson, J. Wrachtrup, B. Zhou, Quantum technologies with optically interfaced solid-state spins, Nat. Photon. 12 516, 2018. 

[3] 	H. J. Kimble, The quantum internet, Nature 453 1023, 2008.\

[4] 	S. Cundiff, A. Weiner, Optical arbitrary waveform generation, Nat. Photon. 4 760, 2010. 

[5] 	D. Awschalom, K. Berggren, H. Bernien, et.al.,, Development of quantum interconnects (QuICs) for next-generation information technologies, PRX Quantum 2 017002, 2021. 

[6] 	F. Kaneda, F. Xu, J. Chapman Kwiat, Quantum-memory-assisted multi-photon generation for efficient quantum information processing, Optica 4 1034, 2017. 

[7] 	R. Weis, T. Gaylord, Lithium niobate: Summary of physical properties and crystal structure, Appl. Phys. A 37 191, 1985. 

[8] 	N. Curtz, R. Thew, C. Simon, N. Gisin, H. Zbinden, Coherent frequency-down-conversion interface for quantum repeaters, Opt. Express 18 22099, 2010. 

[9] 	N. Sinclair, E. Saglamyurek, H. Mallahzadeh, J. Slater, M.George, R. Ricken, M. Hedges, D. Oblak, C. Simon, W. Sohler, W. Tittel, Spectral multiplexing for scaable quantum photonics using an atomic frequency comb quantum memory and feed-forward control, Phys. Reiv. Lett. 113 053603, 2014.

[10] 	J. Lin, F. Bo, Y. Cheng, J. Xu, Advances in on-chip photonic devices based on lithium niobate on insulator, Photonic Res. 8 1910, 2020. 

[11] 	M. Zhang, B. Buscaino, C. Wang, A. Shams-Ansari, C. Reimer, R. Zhu, J. Kahn, M. Loncar, Broadband electro-optic frequency comb generation in a lithium niobate microring resonator, Nature 568 373, 2019. 

[12] 	C. Wang, M. Zhang, X. Chen, M. Bertrand, A. Shams-Ansari, S. Chandrasekhar, P. Winzer, M. Loncar, Integrated lithium niobate electro-optic modulators operating at CMOS-compatible voltages, Nature 562 101, 2018. 

[13] 	M. Li, J. Ling, Y. He, U. Javid, S. Xue, Q. Lin, Lithium niobate photonic-crystal electro-optic modulator, Nat. Commun. 11 4123, 2020. 

[14] 	M. Zhang, C. Wang, R. Cheng, A. Shams-Ansari, M. Loncar, Monolithic ultra-high-Q lithium niobate microring resonator, Optica 4 1536, 2017. 

[15] 	Z. Wang, Z. Fang, Z. Liu, W. Chu, Y. Zhou, J. Zhang, R. Wu, M. Wang, T. Lu, Y. Cheng, On-chip tunable microdisk laser fabricated on Er$^{3+}$-doped lithium niobate on insulator, Opt. Lett. 46 380, 2021. 

[16] 	O. Alibart, V. D'Auria, M. Micheli, F. Doutre, F. Kaiser, L. Labonte, T. Lunghi, E. Picholle, S. Tanzilli, Quantum photonics at telecom wavelengths based on lithium niobate waveguides, J. Opt. 18 104001, 2016. 

[17] 	B. Casabone, C. Deshmukh, S. Liu, D. Serrano, A. Ferrier, T. Hümmer, P. Goldner, D. Hunger, H. de Riedmatten, Dynamic control of Purcell enhanced emission of erbium ions in nanoparticles, arXiv 2001.08532, 2020. 

[18] 	D. Wang, H. Kelkar, D. Martin-Cano, D. Rattenbacher, A. Shkarin, T. Utikal, S. Götzinger, V. Sandoghdar, Turning a molecule into a coherent two-level quantum system, Nat. Phys. 15 483, 2019.

[19] 	D. Najer, I. Söllner, P. Sekatski, V. Dolique, M. Löbl, D. Riedel, R. Schott, S. Starosielec, S. Valentin, A. Wieck, N. Sangouard, A. Ludwig, R. Warburton, A gated quantum dot strongly coupled to an optical microcavity, Nature 622 575, 2019. 

[20] 	N. Sinclair, E. Saglamyurek, M. George, R. Ricken, C. Mela, W. Sohler, W. Tittel, Spectroscopic investigations of a Ti : Tm : LiNbO3 waveguide for photon-echo quantum memory, J. Lumin. 130 1586 , 2010. 

[21] 	S. Dutta, E. Goldschmidt, S. Barik, U. Saha, E. Waks, Integrated photonic platform for rare-earth ions in thin film lithium niobate, Nano Lett. 20 741, 2020. 

[22] 	Z. Kis, G. Mandula, K. Lengyel, I. Hajdara, L. Kovacs, M. Imlau, Homogeneous linewidth measurements of Yb$\mathrm{^{3+}}$ ions in congruent and stoichiometric lithium niobate crystals, Opt. Mater. 37 845, 2014. 

[23] 	F. Bussieres, C. Clausen, A. Tiranov, B. Korzh, V. Verma, S. Nam, F. Marsili, A. Ferrier, P. Goldner, H. Herrmann, C. Silberhorn, W. Sohler, M. Afzelius, N. Gisin, Quantum teleportation from a telecom-wavelength photon to a solid-state quantum memory, Nat. Photon. 8 775, 2014. 

[24] 	M. Zhong, M. Hedges, R. Ahlefeldt, J. Bartholomew, S. Beavan, S. Wittig, J. Longdell, M. Sellars, Optically addressable nuclear spins in a solid with a six-hour coherence time, Nature 177 517, 2015. 

[25] 	E. Saglamyurek, J. Jin, V. Verma, M. Shaw, F. Marsili, S. Nam, D. Oblak, W. Tittel, Quantum storage of entangled telecom-wavelength photons in an erbium-doped optical fibre, Nat. Photon. 9 83, 2015. 

[26] 	R. Kolesov, K. Xia, R. Reuter, R. Stöhr, A. Zappe, J. Meijer, P. Hemmer, J. Wrachtrup, Optical detection of a single rare-earth ion in a crystal, Nat. Commun.  3 1029, 2012. 

[27] 	D. Ding, L. Pereira, J. Bauters, M. Heck, G. Welker, A. Vantomme, J. Bowers, M. Dood, D. Bouwmeester, Multidimensional Purcell effect in an ytterbium-doped ring resonator, Nat. Photon.  10 385, 2016. 

[28] 	A. Dibos, M. Raha, C. Phenicie, J. Thompson, Atomic Source of Single Photons in the Telecom Band, Phys. Rev. Lett. 120 243601, 2018. 

[29] 	T. Zhong, J. Kindem, J. Bartholomew, J. Rochman, I. Craiciu, V. Verma, S. Nam, F. Marsili, M. Shaw, A. Beyer, A. Faraon, Optically Addressing Single Rare-Earth Ions in a Nanophotonic Cavity, Phys. Rev. Lett. 121 183603, 2018. 

[30] 	X. Jiang, D. Pak, A. Nandi, Y. Xuan, M. Hosseini, Rare earth-implanted lithium niobate: Properties and on-chip integration, Appl. Phys. Lett. 115 071104, 2019.

\end{document}